# TAC PROPOSAL FOR FUNDAMENTAL AND APPLIED RESEARCH: LINAC-RING TYPE PHI-FACTORY

Ö. Yavaş, A. K. Çiftçi
Univ. of Ankara, Faculty of Science, Dept. of Physics, Ankara, TURKEY

S. Sultansoy
DESY, Notke Str. 85, D-22607 Hamburg, GERMANY
Univ. of Ankara, Faculty of Science, Dept. of Physics, Ankara, TURKEY
Institute of Physics, Academy of Sciences, H. Cavid Ave. 33, Baku, AZERBAIJAN

*Abstract*
Main parameters of linac-ring type $\phi$-factory proposed as the part of the Turkic Accelerator Complex (TAC) in the framework of ATAM Science City Project are discussed. Two set of parameters, corresponding to E=130 (260) MeV for electron linac and E=2000 (1000) MeV for positron ring, are considered. It is shown that, in principle, L=$10^{34}$cm$^{-2}$s$^{-1}$ can be achieved, which is more than an order exceeds the design luminosity of the DA$\Phi$NE. Parameters of the FEL based on electron linac and synchrotron radiation sources based on the positron ring are estimated.

## 1 INTRODUCTION

The old idea [1] to collide a beam from a linear accelerator with a beam circulating in a storage ring has been renewed recently for two purposes: to achieve the TeV scale in lepton-hadron and photon-hadron collisions (see review [2] and references therein) and to construct high luminosity particle factories. In the last direction linac-ring type B-factory [3], c-$\tau$-factory [4] and $\phi$-factory [5] have been proposed. In this paper we present some preliminary results of the linac-ring type $\phi$-factory studies performed by the Ankara University Accelerator Physics Group [6].

## 2 GENERAL OVERWIEV

The general scheme of proposed complex is given in Fig.1. Electrons accelerated in main linac up to energies 260(130) MeV are forwarded to detector region where they collide with positrons from main ring or turned out to undulator region where FEL beam is produced. On the other side electrons, accelerated in small linac, are forwarded to conversion region where positron beam is produced. Then, positrons are accumulated in booster and after some beam gymnastics are forwarded to the main ring and accelerated up to energies 1(2) GeV. Wigglers installed in two regions will provide SR for applied researches.

## 3 MAIN PARAMETERS OF LINAC-RING TYPE $\phi$ FACTORY

Main parameters of proposed machine are given in Table 1 for two different choices of electron and positron beam energies. Below we present several illuminating notes.

Table I. Main parameters of the $\phi$ factory

| | | |
|---|---|---|
| Electron beam energy, MeV | 130 | 260 |
| Positron beam energy, MeV | 2000 | 1000 |
| Center of mass energy, MeV | 1020 | 1020 |
| Radius of ring, m | 50 | 30 |
| Acceleration gradient, MV/m | 12.5 | 12.5 |
| Length of main linac, m | 11 | 21 |
| Electrons per bunch, $10^{10}$ | 0.04 | 0.02 |
| Positrons per bunch, $10^{10}$ | 10 | 20 |
| Collision frequency f, MHz | 30 | 30 |
| Bunches per ring | 32 | 19 |
| Electron current, mA | 1.92 | 0.96 |
| Positron current, A | 0.96 | 0.48 |
| Energy loss per turn, keV | 30 | 3 |
| Fractional energy loss of the electrons $\delta$, $10^{-4}$ | 2 | 1 |
| Beam size at the collision point $\sigma_{x,y}$, $\mu$m | 1 | 1 |
| Luminosity L, $10^{34}$cm$^{-2}$s$^{-1}$ | 1 | 1 |

Electron bunches accelerated in main linac are used only once for collisions. On the other hand, positron bunches have to be used numerously, therefore, the stability of positron beam is very important. Empirically, allowed values for beam-beam tune shift is $\Delta Q \leq 0.06$ for lepton beams in storage rings. In principle, this upper limit taken from experiments done in usual ring-ring type $e^+e^-$ colliders can be higher for linac-ring type machines. Nevertheless, we use the conservative value $\Delta Q \leq 0.06$ for both options.

The smallness of fractional energy loss of electrons in positron beam field is very important for resonance production of $\phi$-mesons. For this reason $\delta$ should be less than 0.004.

The usage of the flat beams may be advantageous in order to reduce the huge value of disruption parameter for the electrons. The work on the subject is under development.

## 4 PHYSICS SEARCH POTENTIAL

Since deviation of the center-of-mass energy of $e^+e^-$ collisions is smaller than the total decay width of $\phi$-meson, cross-section in the $\phi$ resonance region can be taken as $\sigma \approx 4.4 \cdot 10^{-30} cm^2$.

In the proposed complex $4.4 \cdot 10^{11}$ $\phi$-mesons, $2.2 \cdot 10^{11}$ $K^+K^-$ pairs and $1.5 \cdot 10^{11}$ $K^0K^0$ pairs can be produced in a working year ($10^7$s). Fundamental problems of particle physics such as CP violation, rare decays of K-mesons etc. can be investigated with highest statistics. Moreover, kinematics asymmetry can be advantageous for measuring neutral K-meson's oscillations and CP violation parameters.

Table II. Event numbers in proposed collider

| Decay channels | Branching ratios | LR factory |
|---|---|---|
| $K^+K^-$ | 0.495 | $2.2 \cdot 10^{11}$ |
| $K^0K^0$ | 0.344 | $1.5 \cdot 10^{11}$ |
| $\rho\pi$ | 0.155 | $6.9 \cdot 10^{10}$ |
| $\eta\gamma$ | $1.26 \cdot 10^{-2}$ | $5.6 \cdot 10^9$ |
| $\pi^0\gamma$ | $1.31 \cdot 10^{-3}$ | $5.8 \cdot 10^8$ |
| $e^+e^-$ | $2.99 \cdot 10^{-4}$ | $1.3 \cdot 10^8$ |
| $\mu^+\mu^-$ | $2.5 \cdot 10^{-4}$ | $1.1 \cdot 10^8$ |
| $\eta e^+e^-$ | $1.3 \cdot 10^{-4}$ | $5.8 \cdot 10^7$ |
| $\pi^+\pi^-$ | $8 \cdot 10^{-5}$ | $3.6 \cdot 10^7$ |
| $\eta'(958)\gamma$ | $1.2 \cdot 10^{-4}$ | $5.4 \cdot 10^7$ |
| $\mu^+\mu^-\gamma$ | $2.3 \cdot 10^{-5}$ | $1.0 \cdot 10^7$ |

## 5 SYNCHROTRON RADIATION FACILITY

By inserting wigglers on the straight parts of the main ring of $\phi$-factory, one can produce synchrotron radiation for applied researches. If one use SiCo type magnet: $B_m$=3.33 Tesla, b=5.47 and c=1.8 [7], where $B_m$ is peak value of magnet's field, b and c are constants related to used permanent magnets.

Fig. 2 and Fig. 3 present the spectral flux and central brightness with respect to photon energy for three different values of the vertical distance $g$ between magnets.

Main parameters of SR facility for two options are given in Table III.

Table III. Main parameters of SR facility

| Positron energy, MeV | 1000 | 2000 |
|---|---|---|
| Maximum magnetic field, T | 1.054 | 1.054 |
| Current, A | 0.976 | 0.488 |
| Period, cm | 3.2 | 13.2 |
| Gap $g$, mm | 30 | 30 |
| Total length, m | 2.112 | 2.112 |
| Total radiated power, kW | 1.44 | 2.90 |
| Critical energy, keV | 0.700 | 2.804 |
| Wiggler parameter | 12.99 | 12.99 |
| Spectral flux, Phot/s·mrad·0.1%bandw | $4.96 \cdot 10^{14}$ | $5.00 \cdot 10^{14}$ |
| Spectral central brightness, Phot/s·mrad$^2$·0.1%bandw | $4.95 \cdot 10^{14}$ | $9.98 \cdot 10^{14}$ |

## 6 FREE ELECTRON LASER FACILITY

Free Electron Laser (FEL) is a mechanism to convert some part of the kinetic energy of relativistic electron beam into tunable, highly bright and monochromatic coherent photon beam by using undulators inserted in linear accelerators or synchrotrons [8]. Main parameters of FEL facility for two options are given in Table IV. For using thereafter, magnetic field is estimated to be 1.48 kG with b=5.47, c=1.8, $\lambda_u$=33mm and $g$=25mm. With these values, strength parameter of undulator is K=0.456. Fig. 4 shows the dependence of FEL flux on photon energy for $E_{e^-}$=260MeV option. Here peaks are placed at odd harmonics and maximum values of fluxes are $7.56 \cdot 10^{13}$, $1.08 \cdot 10^{13}$ and $9.45 \cdot 10^{11}$ for n=1, 3 and 5, respectively. Obtained averaged brightness values of photon beam are given in Table IV.

Table IV. Main parameters of FEL facility

| Electron energy, MeV | 130 | 260 |
|---|---|---|
| Photon energy, eV | 4.07 | 16.30 |
| Laser wavelength, °A | 3044 | 761 |
| Current, mA | 1.92 | 0.96 |
| Particle per bunch, $10^{10}$ | 0.04 | 0.02 |
| Repetition frequency, MHz | 30 | 30 |
| Flux, Phot/s·mrad·0.1%bandw | $3.78 \cdot 10^{13}$ | $7.56 \cdot 10^{13}$ |
| Averaged brightness, Phot/s·mrad$^2$·0.1%bandw | $2.91 \cdot 10^{11}$ | $5.81 \cdot 10^{11}$ |

## ACKNOWLEDGEMENTS

This work is supported by Turkish State Planning Organization under the Grant No DPT-97K-120420 and DESY.

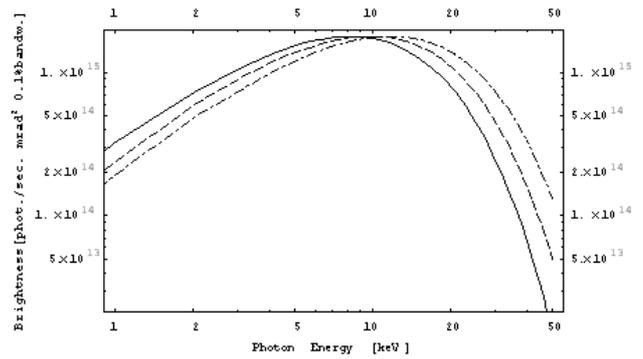

Figure 3. Spectral central brightness of the SR. Solid line corresponds to g=30 mm, dashed line to g=25 mm and dot-dashed line to g=20 mm.

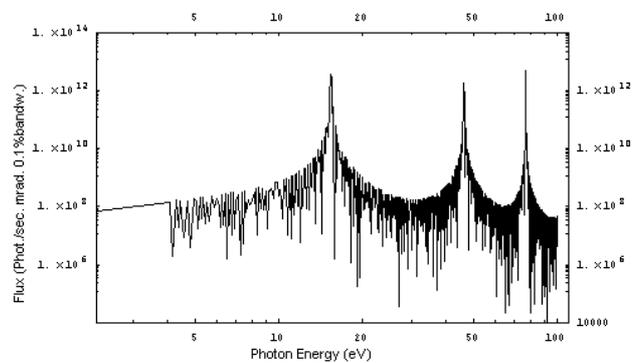

Figure 4. Flux of FEL beam

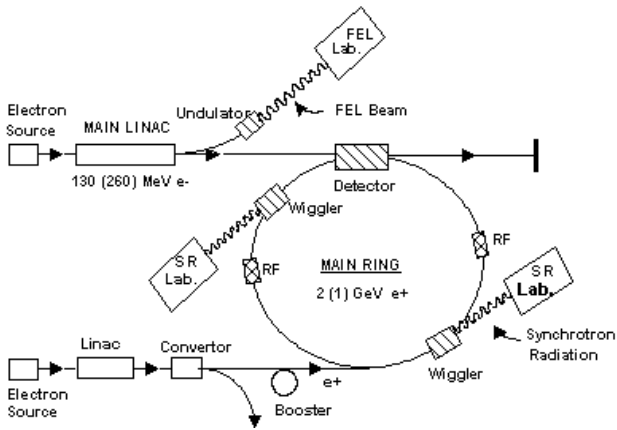

Figure 1. General scheme of the proposed complex

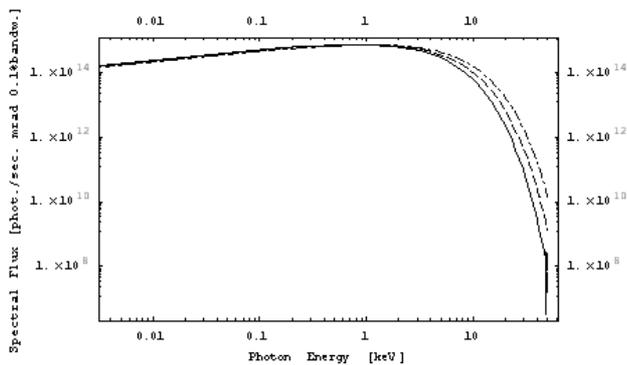

Figure 2. Spectral flux of the SR. Solid line corresponds to g=30 mm, dashed line to g=25 mm and dot-dashed line to g=20 mm.